# Forecasting of COVID- 19 cases using Statistical Models and Ontology-based Semantic Modelling: A real time data analytics approach

SADHANA TIWARI[1], RITESH CHANDRA[2], SONALI AGARWAL[3]
*Indian Institute of Information Technology, Allahabad, India* [1, 2, 3]
rsi2018507@iiita.ac.in[1], rsi2022001@iiita.ac.in[2], sonali@iiita.ac.in[3]

**Abstract**

SARS-COV-19 is the most prominent issue which many countries face today. The frequent changes in infections, recovered and deaths represents the dynamic nature of this pandemic. It is very crucial to predict the spreading rate of this virus for accurate decision making against fighting with the situation of getting infected through the virus, tracking and controlling the virus transmission in the community. We develop a prediction model using statistical time series models such as SARIMA and FBProphet to monitor the daily active, recovered and death cases of COVID-19 accurately. Then with the help of various details across each individual patient (like height, weight, gender etc.), we designed a set of rules using Semantic Web Rule Language and some mathematical models for dealing with COVID-19 infected cases on an individual basis. After combining all the models, a COVID-19 Ontology is developed and performs various queries using SPARQL query on designed Ontology which accumulate the risk factors, provide appropriate diagnosis, precautions and preventive suggestions for COVID Patients. After comparing the performance of SARIMA and FBProphet, it is observed that the SARIMA model performs better in forecasting of COVID cases. On individual basis COVID case prediction, approx. 497 individual samples have been tested and classified into five different levels of COVID classes such as Having COVID, No COVID, High Risk COVID case, Medium to High Risk case, and Control needed case.

## 1. Introduction

Coronavirus (COVID-19) illness is basically a worldwide spreading pandemic, which affects most of the people's lives (in billions) across the world. It is generated by acute respiratory syndrome coronavirus 2 (SARS-CoV-2) virus in 2019. The new SARS-CoV-2 virus was initially introduced in December 2019, Wuhan, China [1]. Plenty of affected countries are taking important measures to set the limit on spread of coronavirus around the world and they have failed yet. On January 30, 2020, a catastrophic situation was announced by the World Health Organization (WHO) and on March 11, 2020 the pandemic became an international level issue [2]. This is the deadliest event in history. Time series data forecasting is necessary for analysing dynamically changing COVID data [2][3]. The technologies related to big data analytics, real time data mining, machine learning (ML), Artificial Intelligence (AI), statistical data modelling and semantic modelling have played a very very essential part in identifying and tackling the situation of COVID-19 emergence [3] [4] [5].

There is an immense need to tackle the spread of SARS-CoV-2 Infectious Disease outbreak by applying data analytics techniques to strengthen the current economic, social and medical emergency situations [2]. The shortage of a consistent framework or prediction model creates a barrier which can be solved with the help of data analytics, machine learning, and semantic web modelling [3][6]. The ontology is helpful to provide a formal definition and description of COVID-19 basics like age group, symptoms, infection rates, contact tracing, drug modelling, and provides a compelling solution to fill this gap [6][7][8]. Many different data sources can be combined through Ontology-based solutions to provide better support for epidemiological situations, and identify pandemic hotspots to support argument, evidence, and knowledge-based recommendations for smart lock solutions [9][10].

To handle the various issues created through SARS-CoV-2, the fusion of statistical time series modelling, semantic modelling through ontologies and big data analytics (BDA) provides an efficient solution for the improvement of the economy, human lives, businesses and occupations across the world [11] [12].The stock market is collapsing day by day as the global financial market is undergoing major changes. However, due to the current epidemic, the Indian economy is slowing down and it is very difficult to recover from it [13]. There are also many significant effects of this infectious disease on the human health systems and social facets of the society. The shortage of diagnostic kits / accessories is also an obstacle to the effective identification and treatment of infected people to control other infections. As a result, the need for sensitive diagnostic tools among healthcare professionals remains important to enable faster identification of potential COVID-19 cases [8][13] .

Researchers are exploring the potential computation models to contribute in this situation of crisis, as current traditional AI models pose challenges, while exploring ways to improve the accuracy of COVID-19 case prediction [12][14]. In this sense, the performance of the traditional machine learning prediction models and big data analytics forms the basis for its use in the detection and prediction of Covid-19 cases [15]. This study investigated an ontology based decision support system integrated with the statistical models for time series data forecast and prediction using big data technology as a real-world model for COVID-19 detection and prediction. The prime contribution of this research is as follows:
- Identify the nature of COVID data.
- Convert the non-stationary COVID data into stationary data using some signification transformation like Log, Subtracting simple Rolling Averages and using the shift() method for accurate prediction.
- To perform accurate forecasts use ARIMA and FBprophet models.
- Compare the performance of both the models using MAE, MAPE and RMSE performance measures.
- Prediction of COVID cases using these models.
- Propose an Ontology based Decision Making System using SWRL rules.
- Incorporation of the proposed mathematical model for precautionary suggestion of COVID patients.

## 2. Background and Related work

On March 11, 2020, WHO declared the novel coronavirus (COVID-19) outbreak a global pandemic [1]. With more than 5 million deaths around the world, the COVID-19wide spread pandemic is considered to be the most serious and critical incident after World War 2. In such unprecedented circumstances, it becomes essential to understand the growth and behaviour of the pandemic [1][2]. Many prediction methods have been developed in the literature for the detection and prediction of active case, deaths and recovered cases from the COVID-19 dataset. These prediction models are also helpful in predicting the future cases in any particular country or region [3].

### 2.1 Time series data forecasting using Statistical models

One of the most prominent methods used in time series prediction is the Auto Regressive Integrated Moving Average (ARIMA) model [4] [5]. This model allows the customization of time series data for prediction of future occurring data of the series. Seasonal ARIMA (SARIMA) is another model for time series data prediction including seasonality [16]. Statistical Time series models are helpful for modelling and predicting time-indexed data. ARIMA model can be broken into three main terms as AR, I, MA: where,

AR (p) is used for an autoregressive model by passing parameter p which represents an integer and which certifies the number of lagged series used for future predictions.
I (d) represents the differential part and parameter d informs about differencing orders which are being applied for constructing the series stationary.
MA (q) is used for moving average where parameter q tells about the terms of total lagged forecast error in prediction of time indexed data.

By carrying out a pragmatic analysis of the dataset, the identification of a suitable model is necessary, that could be readily used by researchers, medical experts, organizations, society, and governments to investigate the recent future outbreak of this event. The logic behind the selection of ARIMA and SARIMA time series model are as follows:
- i) These models provide very satisfactory results in predicting natural adversity compared to other predictive models like the Wavelet Neural Network (WNN) model and the Support Vector Machine (SVM) model.
- ii) The statistical models have been used before in similar crisis situations, such as during the SARS epidemic. This model is used to predict the number of beds occupied by the Tan Tock Bed Hospital in Singapore in real time [5]. Furthermore, Anwar et al. Developed a tool to predict malaria patterns in

Afghanistan using the ARIMA model [16]. Later, Benevento et al. have used this method to generate the upcoming predictions of COVID-19 worldwide [17].

## 2.2 Time series data forecasting through Facebook Prophet (FB-Prophet) models

Facebook Prophet is an additive model-based time series forecasting method that uses nonlinear trends to determine daily, weekly, monthly, and yearly inclination, and also focuses on holiday effects. The FB-Prophet model is resilient to various data pre-processing issues like missing value, repeated values and trend changes and generally handles outliers very well [18]. This model can be trained just like a curve fitting practice and ignores the time sensitive part of the data. Therefore, improper observation may be permitted in the dataset [19]. The various benefits of applying this model on time series data, are: Offers multi-period seasonality; Suitable for known and customized holidays; Offers flexibility with two popular options:
1. Linear model
2. Saturation growth model which can adapt changes very quickly. [20]
This predictive modelling technique is beneficial to rename the Date and the output column to predict future cases. Furthermore, the Date should be converted to Date Time format.

## 2.3 Combining Statistical data models with Semantic data modelling using Ontologies

Statistical data models, machine learning and semantic data modelling using ontologies are the most commonly used technologies helping behaviour, trends and data prediction. Although the performance of these models is not limited to the data availability but also depends on the quality of input data [7] [11]. For efficient data processing structured information is needed for machine learning and training to achieve good prediction accuracy of Covid-19 dataset. So, transformation of unstructured data into meaningful information is one of the immense challenges. For this purpose computational ontologies are helping a lot for handling unstructured, completely insufficient and heterogeneous data. Earlier Machine learning, statistical data modelling and semantic modelling through ontologies are considered as different approaches, but nowadays, fusion of these technologies are becoming popular due to huge amounts of data and increasing complexity [7]. Table 1 presents the summarized literature review in the area of predictive modelling and Semantic web modelling with the help of ontologies and machine learning.

Table 1: Fusion of predictive data analytics and Semantic (ontology) modelling in COVID-19 scenario

| Paper ID | Problem | Proposed solution | Dataset/use case used | Tool type | Outcome |
|---|---|---|---|---|---|
| [8] 2021 | Early detection of COVID-19 patients having high level of complication | Predict critically of any patients within 28 days of diagnosis including symptoms, isolation, demographics, treatment, comorbidities and hospitalization | US electronic health records (IBM Explorys) | Novel Web Platform | Various performance metrics ROC AUC, PR AUC, Brier score, Log loss, Sensitivity, Specificity, and F1-score have been computed. |
| [6] 2020 | Prediction of COVID -19 pandemic cases | A novel solution is proposed which combines AI, Big data and Semantic web services (SWS) | Coronavirus prediction case study | Protege (Standalone Platform) | Detection of suspected patients of COVID-19 to accurately identify the spreading rate. |
| [7] 2021 | Monitoring physiological parameters of students during COVID-19 | An ontological framework CCOnto to describe situational behaviour in humans | COVID-19 use case | COVID19 dashboard | Analysing change in behaviour of university students due to COVID pandemic through ontological rules. |
| [9] 2020 | Critically analysing interconnection among hosts and different variants of coronaviruses | Ontology based solution is proposed and classification of host-coronavirus interactions (HCI) and disease prediction | 35 tested and originally collected HCI protein-protein interactions (PPIs) | Python | Proposed logical ontology with computational prediction model develops understanding of human patients of |

| | | | | | COVID-19 and investigate their correlation with other diseases |
|---|---|---|---|---|---|
| [10] 2020 | Exhaustive labour cost in data preparation, involvement of non-suitable experts in manufacturing analysis and shortage of generalized ML models | Semantically improved ML pipeline combined with feature engineering (SemFE) provides a bridge to fill the gaps in this area and makes data science available for all non ML experts. | COVID-19 scenario | Protege, Onto-NoSQL | Applying semantic data pre-processing and semantic feature engineering with ML pipelines to forecast the infected cases of COVID-19 |
| [11] 2021 | Sentiment analysis and prominent topic exploration of COVID-19 tweets | Natural language processing methods along with machine learning models are used for analysis of COVID-19 tweets | COVID-19 twitter dataset from IEEE data port | NA | A clustering based classification and topics modelling method (TClustVID) is proposed and model performance has been evaluated |
| [12] 2020 | Handling the big data challenges in case of COVID-19, event tracking and reasoning | Develops an ontology based solution Onto-NoSql using Protege and extracts big data on semantic platforms. | Case studies on weather and air pollution and weather and COVID-19 | Protege, Onto-NoSQL | Predict COVID-19 ubiquity and interconnection with weather, attains 96.9% accuracy, able to handle time and size related challenges. |
| [14] 2020 | Detection of COVID-19 disease in beginning and diagnose as per criticality level | Case based reasoning (CBR) model is used and a new mathematical model based on Semantic modelling is proposed. | COVID-19 dataset | Protege (Standalone Platform) | Accuracy of suspected is achieved as 94.54% |
| [15] 2021 | Handle the challenges of community spread of COVID-19 infections, reduce the risk factors in the highly dense areas | Cyber physical systems and machine learning are helping in monitoring the spreading risk of coronavirus in heavily populated areas. | COVID-19 dataset | Protege, RDF-Gen. | Various performance measures have been computed such as accuracy, f-measure, execution time, kappa, MAE, RMSE etc. |
| [21] 2021 | Issues related to sensory data generation, interoperability of IoT devices, and considerable social and environmental issues | Framework designed which integrates Internet of Things (IoT), machine learning and other AI based solutions for handling COVID-19 in smart cities. | COVID-19 case study of New York, China, Canada and USA | NA | Handle the COVID-19 cases and help the corona warriors to fight against this pandemic |
| [22] | Integrating COVID-19 open data collected | An ontology-based technique for acquiring, representing, and | US electronic health records | Novel Web Platform | COVID-19 Ontology Submitted in OBO Foundry. Open source |

| | from various sources to make decisions. | transforming data to RDF, that allows data to be linked with data from other publications freely accessible data sources | (IBM Explorys) | | for Decision Making for COVID-19 patients. |

After studying various research in COVID-19 prediction, it is identified that there is significant scope for work in forecasting and tracking the situation of COVID-19. Applying different machine learning models, statistical forecasting approaches [23][24] and semantic web modelling through ontology [9][13], seems a novel interdisciplinary idea for COVID-19 prediction, prevention and treatment. In such unprecedented circumstances, it becomes essential to understand the growth and behaviour of the pandemic. This research is broadly classified as follows:
(1) Application of machine learning, big data analytics and statistical modelling to identify patterns, nature and trends of several time variant events related to various infectious diseases.
(2) Latest research techniques such ontology based modelling is used, which strictly focuses on predicting COVID-19 outbreak-related statistics such as active cases, deaths, recovered cases, chances of getting infected on behalf of body symptoms and chances of getting recovered using statistical models and semantic models.

### 3. Proposed Methodology

### 3.1 Dataset Description

In present research, the Statistical data published by Our World In Data (OWID) Organization [25] is obtained from kaggle [26]. The Data is available in CSV file format and includes information regarding new cases, total cases, and total deaths. Total number of columns in the dataset are 10 named as date, location, new_cases, new_deaths, total_cases, total_deaths, weekly_cases, weekly_deaths, biweekly_cases, biweekly_deaths and size of dataset is approx 4 MB.

### 3.2 Structure of the proposed model

The primary goal of the predictive analytics and decision support system is to detect severe instances of COVID-19. The proposed decision support has several components as shown in figure 1 and it is utilizing collective knowledge provided by applying various rule editing tools, data mining concepts, statistical data modelling, big data analytics, machine learning, AI and ontology based modelling. These methodologies are helping in both diagnosis and prognosis. It takes account of a variety of body vital parameters and symptoms related to COVID-19. The capacity of human biological responses to combat with the virus is a subject of study in this model. Patients with severe COVID-19 infections are diagnosed by general practitioners, specialists, nurses, and physicians with the use of clinical decision-support technologies.

In this work, a four layer architecture has been proposed as shown in figure 1. The first layer is responsible for data collection, where data is gathered from various publicly available repositories and from existing biomedical literature. The second layer is responsible for prediction and Ontology development. The total no. of active, deaths recovered cases is predicted using SARIMA and FBProphet model. For ontology development and defining all the precautionary measures the medical literature available on COVID-19 is used. Then we add the prediction model in Ontology. The COVID-19 dataset is also being analysed using ontologies, followed by ontology-based queries processing. Data annotation and storage is the third layer of the architecture where we added the individual in the ontology in which the COVID cases are categorized into five different labels for individual basis COVID case prediction and the training should be properly annotated as per COVID use case. Store the data into Resource description Format (RDF) using COVID-19 Ontology. In the last layer, using SPARQL to query the ontology for different case situations on behalf of to identify COVID positive and negative cases on the basis of some essential symptoms, as well as other critical symptoms and visualize.

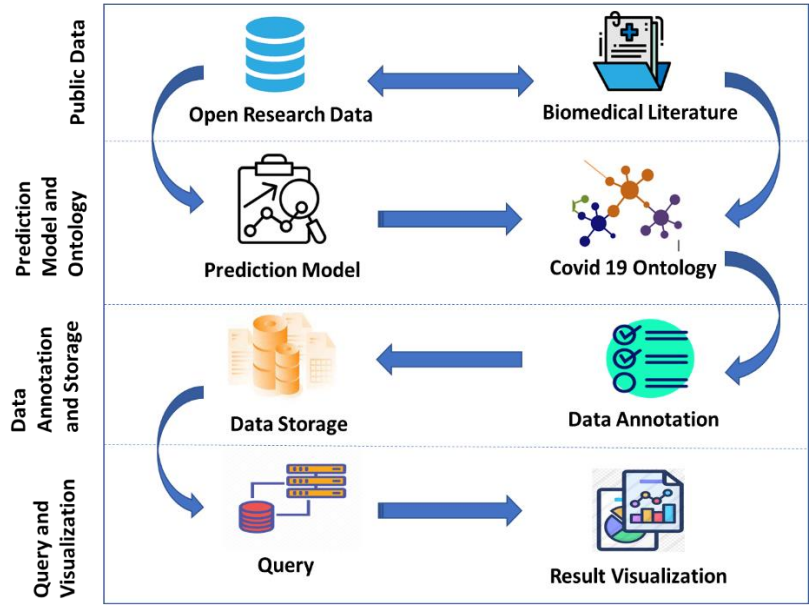

Figure 1. Architecture for fusion of Prediction Model to develop Ontology based decision support system

In this paper, we perform structural analysis of SARS-CoV data through statistical data modelling and predicting the COVID cases through SARIMA and FBprophet models. The COVID-19 dataset is also being analysed using ontologies, followed by ontology-based queries processing to identify COVID positive and negative instances on the basis of some essential symptoms, as well as other critical symptoms. Predictive data analytics is basically a part of advanced data analytics through which future predictions can be made with the help of existing historical data and also identify the data patterns to figure out the future trends, scope, and risks. This stream data mining can be visualized as a combination of statistical data modelling, data analysis with data mining concepts and machine learning methods [12][15]. Many data analytics tools, such as Hadoop, Storm, Spark, Mahout, Drill, and SCALATION, are available to support the analysis of large datasets. In the proposed model, statistical data model ARIMA and FB-Prophet and semantic modelling technology (ontology) are used for selecting, constructing, and explaining the prediction of daily active cases, new cases, and total death cases. Statistical models can be properly explained with the help of analytics ontology [7][15]. Figure 2 shows the primary flow diagram of the proposed model.

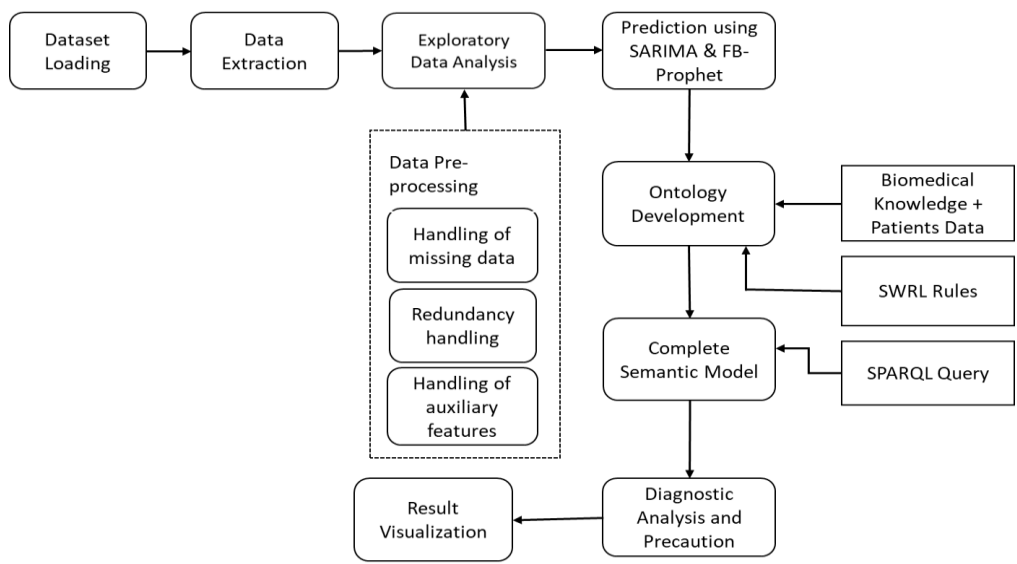

Figure 2: Workflow of the proposed model

The stepwise processing the proposed predictive analytics model is as follows:

Step I- There are two parts to this first step: loading COVID-19 data and making a process that can be used to manipulate data for simple data pre-processing.

Step II- The data extraction step helps in the extraction of key information from the dataset using data and metadata. This step is very essential for the model type selection as well as the validation step.

Step III- This step is useful for data refinement and preparation of the data for analysis purposes. This part includes the handling of missing values, handling of duplicate values and identification of auxiliary features of the dataset.

Step IV- Exploratory data analysis step is essential for understanding the trends and patterns present in the COVID-19 dataset and discovering the anomalies with the help of graphical visualization and statistical summaries.

Step V- The selection of an appropriate model type depends on the dataset characteristics. It is performed by applying domain description across each column of the dataset, and after that, applying predictive analysis using SARIMA and FB-Prophet models to identify the promising model for the respective dataset.

Step VI- This step is needed for validation of the selected model and to move back if the model is not performing well as per the requirements and expectations, then the model needs to be refined.

Step VII- The performance of the various models will be compared with respect to various performance metrics, and the best suitable model will be selected.

Step VIII- In ontology based semantic modelling, some important features will be selected from the dataset and other factors will be associated, such as height, weight, age, gender, and prior illness from some other chronic disease for negative and positive COVID case prediction.

Step IX- After comparison, when the best model is selected and validated, and the model is used for prediction of COVID-19 cases, the COVID-19 ontology generates inference rules using Semantic Web Rule Language (SWRL) to cover all queries related to COVID-19 case prediction, including precaution and chances of having a positive or negative RT-PCR.

## 3.3 Model description

The proposed model employs a statistical time series approach to forecast the coronavirus cases in India for the next 6 months during the first wave of COVID-19. Time series data analysis consists of a series of data that arrives (is indexed or graphed) in a timely manner. Thus, the data must be arranged with moderately deterministic timestamps and can be compared with random samples having some supplementary information. The objective of time series data analysis can be conveyed in two ways: the first way figure out a model which processes an observed sequence of real time data using stochastic mechanisms and the second one forecasts the upcoming values of that sequence according to the past records of the data [27], as shown in figure 3.

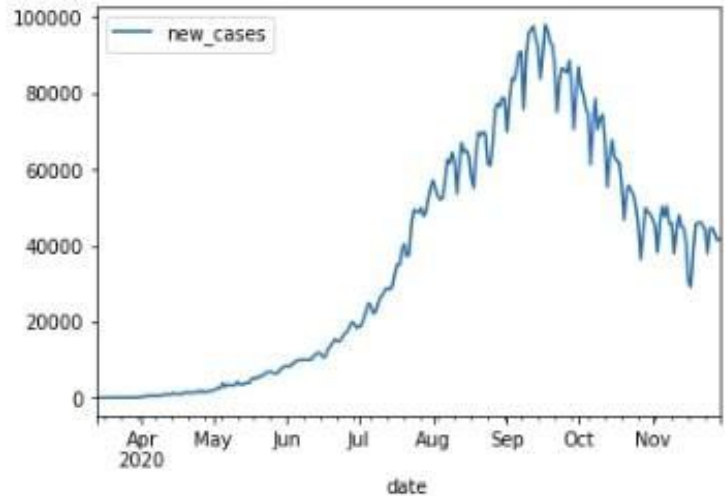

Figure 3: Time series Analysis of daily new cases in India

The analysis of time series data adds an explicit order dependence between observations. It adds a time dimension, which is both a constraint and a structure that provides a source of additional information.

### 3.3.1 ARIMA and SARIMA Modelling for COVID-19 prediction

The Seasonal Auto-Regressive Integrated Moving Average (SARIMA) model is a class of models that explain a given time series dataset based on its lags and the lag forecast errors. The initial steps taken in this work deal with performing exploratory data analysis on the COVID-19 dataset so that we can familiarize ourselves with the underlying trends and patterns present in the data [5]. During this process, the missing values present in the data were removed. Another important criterion that serves as the basis for all ARIMA models is that the data should be stationary. A series is said to be stationary if the marginal distribution of the output Y at time t is the same as at any other point in time. In other words, basic statistics such as mean and variance are time-invariant. One popular way of detecting stationarity in the data is by using the Augmented Dickey-Fuller (ADF) test. The ADF test is a statistical test that returns a set of parameters. If the p-value is less than 0.05, then the data is considered to be stationary [15] [16].

```
Results of Dicky Fuller:
Test Statistic                  -1.866733
p-value                          0.347869
#Lags Used                      15.000000
Number of Observations Used    246.000000
Critical Values (1%)            -3.457215
Critical Values (5%)            -2.873362
Critical Values (10%)           -2.573070
dtype: float64
```

**Figure 4**: The initial results of the ADF test

From the results shown in figure 4, it can be observed that the data is non-stationary since the p-value crosses the 0.05 mark. Non-stationary data can be converted into stationary data through a sequence of transformations applied to the time series data. These transformations are as follows:
1) Taking Log
2) Subtracting simple Rolling Averages
3) Subtracting Exponential Rolling Averages
4) Subtracting Previous Values using shift()
5) Seasonal Decomposition

6) Combination of all

The proposed work uses three significant transformations, the Log, Subtracting simple Rolling Averages and using the shift() method for forecasting COVID-19 using statistical models as given in figure 5 [16][17].

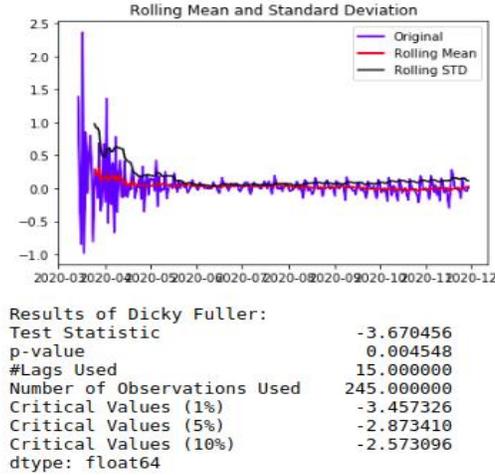

Figure 5: Final results of ADF Test

The ARIMA model can be applied by using three main factors, the moving averages (q), the auto regressing lags (p), and the integrating factor (d). So, it is required to find the optimal values of p, d, q to find the best fitting ARIMA model. These values can either be determined using the ACF, PACF plots given in figure 6 or the auto ARIMA function.

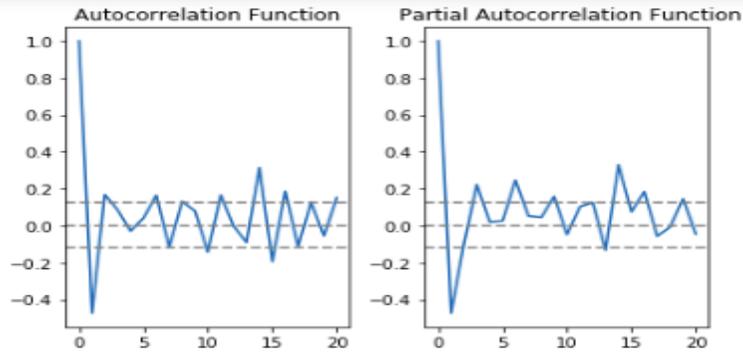

Figure 6: ACF and PACF Plots

Algorithm 1 specifies the sequence of steps applied in predicting COVID-19 infected cases using ARIMA and SARIMA models.

Algorithm -I

1) Get the list of source data files in the specified directory.
2) Reading the CSV file using the read_csv function.
3) Extracting India's Information:
    country1=df.loc[(df['location']=='India'),:]
4) Performing ADF Test to check for stationarity of data.
5) If data! = stationary:
    Making data stationary using above methods
6) Determine ACF and PACF plots for getting optimal values for p, d, q.
6) Divide data in train and test datasets for getting predictions.
7) The arima_model function is used to determine the appropriate values for p,q,d.
8) The predict() of function is used to forecast the cases for the next nine months

### 3.3.2 FB-Prophet model for COVID-19 prediction

Another model used in this work is FB-Prophet for prediction, first define and configure a Prophet () object, then call the fit () function to pass the data to fit the dataset. The argument is passed to the Prophet () object to configure the required model type, such as growth type, seasonality type, and so on. The input of the fit () function is a Data Frame of COVID-19 dataset, which requires a specific format. The first column of the dataset must contain the date and time. The next column must contain the observations. This means renaming the columns in the dataset [18] and [20]. Also, if the first column hasn't been converted to a date time object yet, it needs to be converted. Further analysis of COVID-19 data is obtained by considering the FB Prophet model in the Spark Environment. The COVID-19 dataset obtained from Kaggle [26] is analysed and the output is streamed using Spark Context [28] [29].

### 3.3.3. Semantic modelling through Analytics ontology for COVID-19 prediction

Scientific communities are increasingly looking to ontologies to facilitate web-based management and interchange of scientific data, thanks to recent improvements in data modelling and increased usage of the Semantic Web. Ontologies can be used inside a domain to formally express ideas and links between concepts. The resulting logic-based representations provide a conceptual model that can help with data storage, management, and sharing among numerous research groups. In this paper developed COVID-19 ontology is developed, which improves the accuracy of prediction models and aids decision-making as shown in Figure 7.

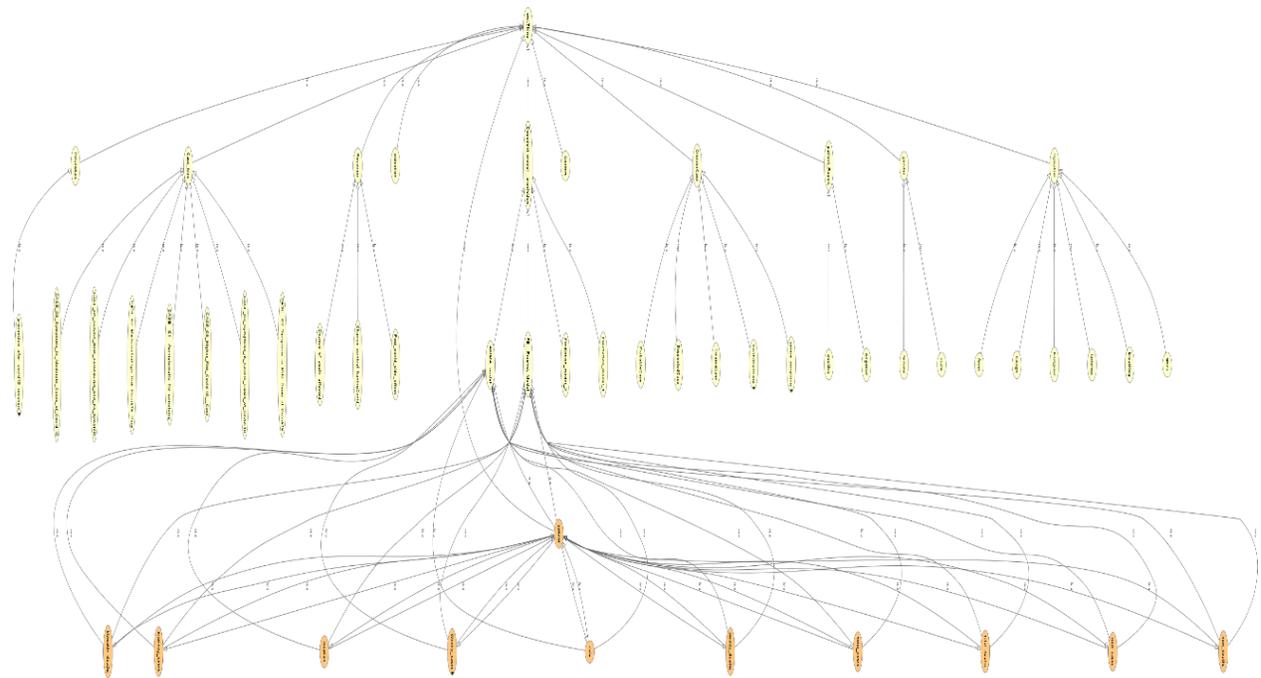

Figure 7. Owl Viz (https://protegewiki.stanford.edu/wiki/File:OwlViz.jpg) view of Hierarchy of COVID-19 ontology

a) **Ontology Design phase**

An ontology is built via a seven-step process: (1) Determine the domain and scope of the ontology. (2) Consider repurposing existing ontologies (3) A list of significant concepts in Ontology (core classes) (4) Classify the classes and their hierarchies. (5) Slots are used to define the properties of classes (6) The facets of the slots must be defined and (7) Instances are created.

b) **Covid-19 Core Classes and Property**

The Covid-19 Ontology, an open source software for creating ontologies, is being developed with the Protégé editor [30]. Several concepts, attributes, and individuals are included in this Covid-19 Ontology, which is based on COVID-

19 literature data, Daily medical conceptual data, prescription suggestions, and key metrics and parameters connected to Covid-19 statistical prediction. An ontology starts from the "Owl:Thing" class, which will be broken down into sub-concepts like "case type," "Dependent model prediction," "Disease Cases," "Gender," "Patient report," "Recovery," and so on. Figure 8 represents the class hierarchy in which numerous Object Properties are included such as "has_Symptom," "hasGender," "has_Symptom_Severity," "hasSuggestion", "has_joint_pain," "has_report" and so on. Further, many Data Properties are added such as "has_asthma", "has_depression", "has_age", "has_type1Diabeties", has_weak_immune", "has_type2Diabeties", "is Asymptomatic" and so on. The "OWL individual" declarations are another semantically ordered relation within the classes. The "Ideas" notation within an COVID-19 Ontology implies a range of OWL individual suggestions, such as SUG 1, SUG 2, SUG 3, SUG 4, SUG 5, and so on, that assist the patient in taking precautions in accordance with this proposal. Patient information with several "cases" concepts denote a variety of case types represented by OWL individuals, like CASE 01, CASE 02, CASE 03 etc.

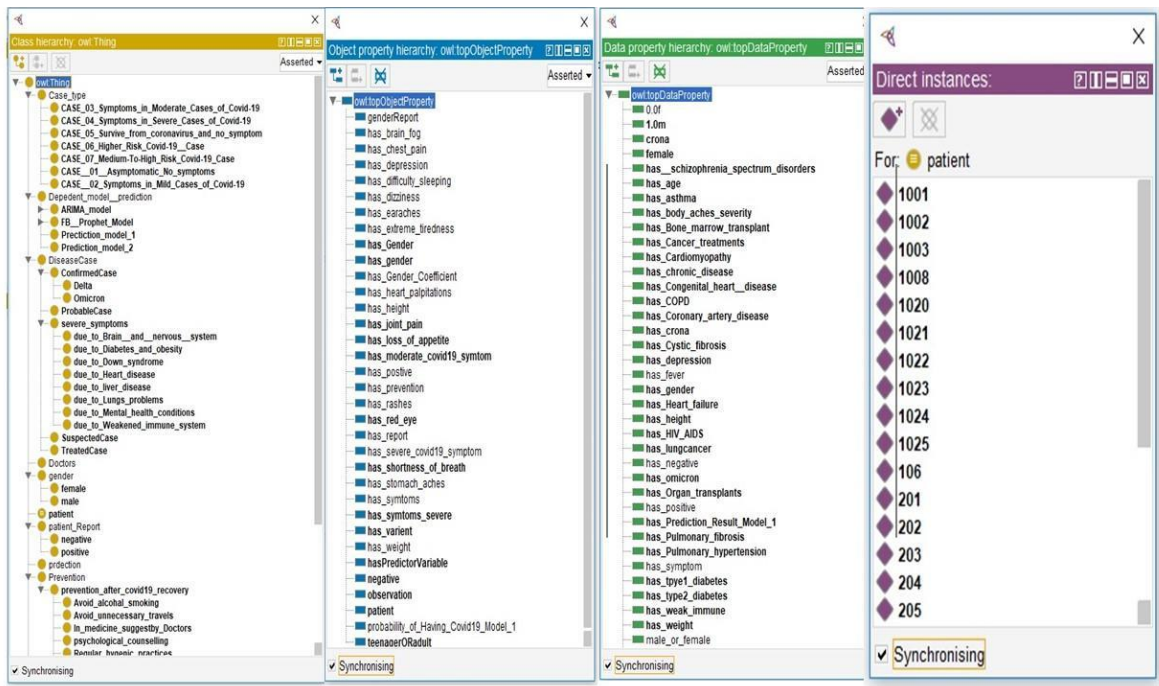

Figure 8. Core Classes, Data Property, Object Property and Individuals (Instances) of COVID-19 Ontology

Technical specifics concerning semantic rules and inferencing will be discussed in the next section. The system that follows and uses those concepts to produce a prediction about the chances of getting affected by COVID-19, as well as individualised supportive action suggestions will be provided based on the criticality of the case.

For making the ontology we took the data from the official website of the Ministry of Health and Family Welfare [REF website https://www.mohfw.gov.in/]. This dataset is collected by the researchers of Indian Statistical Institute [ref: https://www.isibang.ac.in/~athreya/incovid19/data.html] also used SARS-CoV-2 data [REF: https://www.ncbi.nlm.nih.gov/sars-cov-2/]. The Cellfie Protégé plugin [31] is used to reintegrate this data into ontology. Through Cellfie we easily convert spreadsheet rows into instances of class and property. Each row has a specific patient, details of COVID, weekly death cases, weekly infected cases, total cases etc. Table 2 shows the transformation rules of how the dataset is converted into an ontology.

Table 2: Transformation rule

```
Individuals: @B*(mm:hashEncode rdfs:label=("Patient", @B*))
Types: Patient
Facts: 'diagnosed on' @B*(xsd:dateTime),
'Age' @C*(xsd:decimal),
'Has gender' @D*,
'State' @F*,
'City' @E*,
'Traveled from' @G*,
'Nationality' @I*,
'Weeklydeath'@W*
'Location'@l*,
'Toatlcases'@T*,
'Newdeaths'@D*,
'Biweeklydeaths@BW*,
'Status' @J*,
'Deaths'@Z*,
'has resulted in any other infections' @L*(xsd:boolean)
```

### c) The Reasoning Rules for COVID-19 Ontology

The Semantic SWRL [32] is a programming language which is used for defining semantic rules to provide improved provable reasoning capacity. SWRLs are being used to build and validate semantic rules that provide the user with a mix of issue definition facts and knowledge base inference. For designing this ontology, the SWRL rules and OWL axioms use the ROWL plugin [33], available on the top portion of Protege's SWRL user interface. This interface gives the facility to supply input in the form of semantic rules. Any user can generate a rule by applying the SWRL syntax available in the ROWL tab. When any one uses the ROWL plugin to add a rule to the ontology and converts it to OWL axioms, also ROWL increases the amount of OWL axioms.

For Covid -19 Ontology, 43 SWRL rules have been designed based on four main criteria: (1) Mathematical statements, (2) Considerate having COVID-19 through probability, (3) Precautionary guidelines and tailored supporting action recommendations, and (4) SWRL rules for predicting COVID-19 Cases using FBprophet and ARIMA model. Various parameters related to any person, such as weight, height, age, gender, temperature, and some newly observed symptoms (as per their severity levels, likewise, severe, moderate, and mild) are used as inputs in the majority of the rules. Furthermore, the construction of the rules includes various parameters of symptom categories from the mathematical formulation used. By using the patient data and symptom factors, a prediction score is calculated for the likelihood of the chances of having COVID-19. As per the severity level, the COVID infected patients have been divided into five categories: (1) Asymptomatic cases, (2) Severe COVID symptoms, (3) Moderate COVID symptoms, (4) Healed Covid symptoms, (5) and Mild Covid symptoms. The category to which the case belongs is selected based on the score, and a selection of specific supportive action options designated for that category is offered. Following the calculation of the chance of obtaining Covid-19 scores based on statistical models, the suggestion is given according to the score of case type classification, which is mentioned in figure 9. Each case has a different precaution and suggestion linked to it. A total of seven cases are included in our ontology.

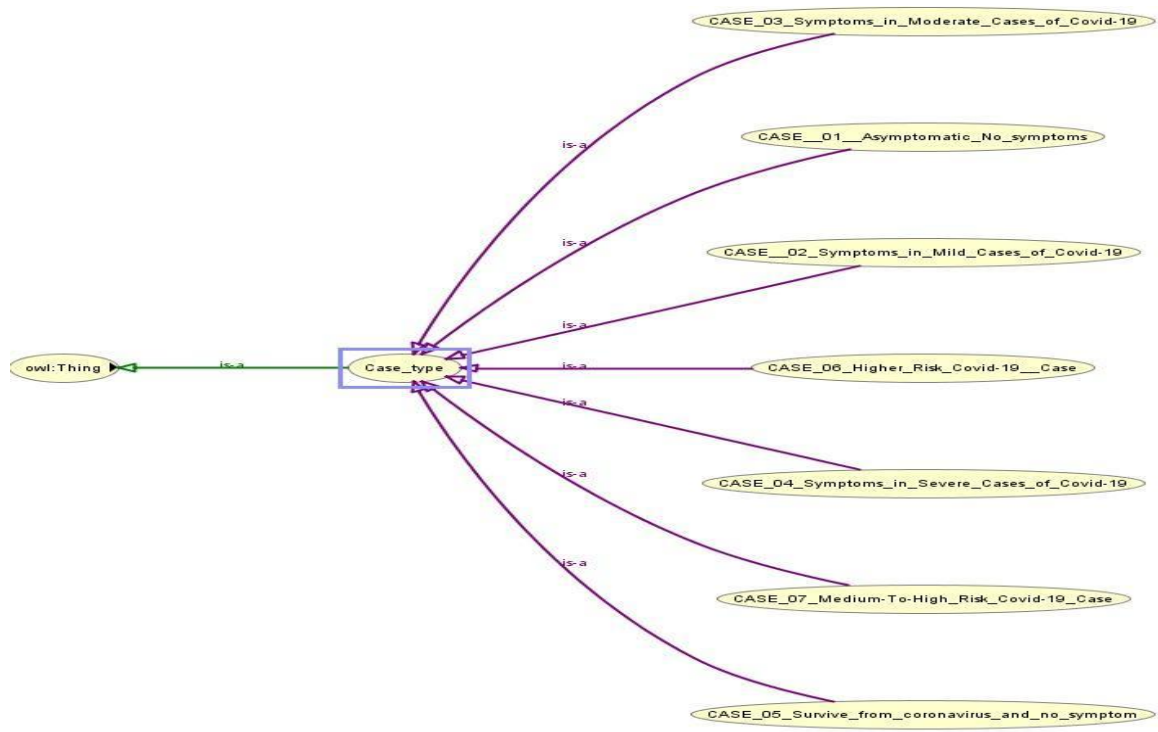

Figure 9. Hierarchy of the Case type

**d) SWRL rules for Mathematical statements**

The complex mathematical formulation can be generated with the help of SWRL rules applying SWRL mathematical expressions resident library [34]. A rule for computing Body Mass Index (BMI) using the SWRL's built-in function is given below:

SWRL Rule of Body Mass Index: patient(?p)^has_weight(?p,?w)^has_height(?p,?h)swrlm:eval(?BMI, "703*w/(pow[h, 2])",? w,? h) - > hasBMI(?p,? BMI).

Where, "? p" represents patient information, "?h", "?w", "?BMI" represents height, weight, and BMI values, respectively. The "swrlm:eval" function determines the value of the "?BMI" variable. Here the symbol "^" is used as a "AND" in logical operators. Input is taken as height and weight for solving BMI of the patient.

**e) Probability Calculation Rules having COVID-19**

The first rule is based on the patient attribute age, sex, loss of smell, loss of flavour, cough, severe fatigue, and missed meals. The second rule is based on the patient attribute age, sex, loss of smell, loss of taste, cough, severe fatigue, and skipped meals.

Prediction Model 1 in [35] = - 1.30 - (0.01 * age) + (0.44 * sex) + (1.75 * loss of smell and taste) + (0.31 * severe persistent cough) + (0.49 * severe fatigue) + (0.39 * skipped meals)   ---(1)

In equation 1, if symptoms are yes then '1', if no means '0'. Gender represented '1' as male and '0' as female. As a result of solving equation 1, this value will be translated to the expected probability using the formula exp(x)/ (1+exp (x)). Estimated COVID-19 cases are assigned odds more than 0.5, whereas controls are assigned probabilities less than 0.5. All SWRL rules for equation 1 are correlated to each other which is shown in figure 10.

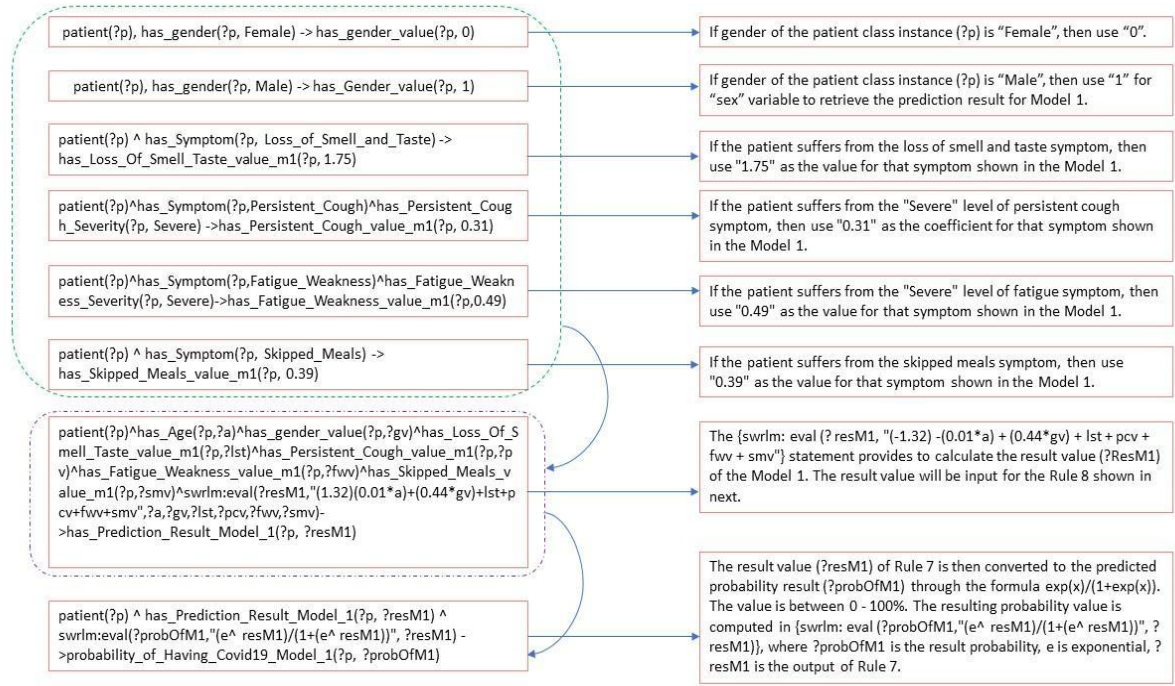

Figure 10. SWRL rules for Model 1 and its working process

In figure 10, all six rules output is input for the second last rule and output of the second last rule is the input for the last rule which gives the output for equation 1.

In Prediction model 2, 10 symptoms have been considered: chest pain, hoarse voice, abdominal pain, skipped meals, and delirium, and diarrhoea, shortness of breath, fatigue, persistent cough, and fever. Here, both models are based on Logistic regression.

Prediction Model 2 in [36] = - 2.30 + (0.01 * age) - (0.24 * sex) + (1.6 * loss of smell stage) + (0.76 * fever) + (0.33 * persistent cough) + (0.25 * fatigue) + (0.31 * diarrhoea) + (0.46 * skipped meals) - (0.48 + abdominal pain) — (2)

As a result, precise SWRL rules are derived from equation 1 and equation 2 to estimate the probable chances of getting a COVID-19 score. It is further categorized in 3 different classes based on prediction score (1) "Higher Risk COVID-19 Case", if probability value is greater than 0.85. (2) "Medium-To-High Risk COVID-19 Case", if probability in between 0.5 and 0.85. (3) "Control-Needed Case", if probability less than 0.5. These all classes give risk factors for COVID-19 patients.

**f) Rules for Precautionary guidelines and tailored supporting action recommendations**

Some Rules are divided into five categories (1) Asymptomatic Case, (2) Severe COVID symptoms, (3) Moderate COVID symptoms, (4) Healed COVID symptoms, (5) and Mild COVID symptoms. Each category has different rules and parameters which define the case type (which is shown in Figure 12) and precautionary steps (https://www.who.int/emergencies/diseases/novel-coronavirus-2019/advice-for-public).

Table 3. SWRL rules for COVID19 different case situation

| Covid 19 different case situation | SWRL rules for Covid 19 case |
|---|---|
| Asymptomatic case | patient(?p)^has_symptom(?p,No_symptom)^hasContactWithCovid19Patient(?p,true)^hasTravelledToCovid19Area(?p,false)→Asymptomatic(?p)^is_diagnosed_with(?p,Suspected_asymptomatic_case)^is_recommended_with(?p,Suspected_asymptomatic_case recomandation) |
| Non suspected case | patient(?p)^has_symptom(?p,No_symptom)^hasContactWithCovid19Patient(?p,false)^hasTravelledToCovid19Area(?p,false)→Nonsuspected_case(?p)^is_diagnosed_with(?p,Nonsuspected_case)^is_recommended_with(?p,Nonsuspected_case_recommendation) |
| Suspected Symptomatic case | patient(?p)^has_symptom(?p,symtom)→Symptomatic(?p)^is_diagnosed_with(?p,Suspected_symptomatic_case)^is_recommended_with(?p,Suspected_symptomatic_case_recommendation) |
| Probable case | patient(?p)^has_symptom(?p,Breating_Difficulty)→Probable_case(?p)^is_diagnosed_with(?p,Probable_case)^is_recommended_with(?p,Probable_case_recommendation) |
| Confirmed case | patient(?p)^has_lab_test(?p,RTPCR)^hasLabTestRTPCRValue(?p,true)→Confirmed_case(?p) ^ is_diagnosed_with(?p,Confirmed_case) ^ is_recommended_with (?p,Confirmed_case_recommendation) |
| Self Quarantine No Medication | patients(?p)^has_symptoms_of(?p,no_symptoms)^ContactWithCOVID-19Patients(?p, false) ->SelfQuarantineNoMedication(?p) |
| Intensive Care With Medication | Patients(?p)^has_symptoms_of(?p,serious_symptoms)^has_risk_of(?p,high_risk)^has_medical_history_of(?p,chronic_diseases)->IntensiveCareWithMedication(?p) |
| Self Quarantine With Medication | Patients(?p)^has_symptoms_of(?p,common_symptoms)^has_risk_of(?p,low_risk)^has_medical_history_of(?p,non_chronic_diseases)→SelfQuarantineWithMedication(?p) |

In Table 3, some other case situation rules of COVID-19 like "Self Quarantine No Medication", "Intensive Case with Medication", etc. are listed. In Figure 11, we demonstrate how the Pellet Reasoner [37] works in Protege finds CASE 03, which comes under the Moderate COVID Symptoms case category. Patient number 101 is male and adult with mild symptoms. What precautionary steps have to be taken, also provided by inference rules. A total of 1250 patient details are included in the COVID-19 ontology and 43 SWRL rules. The pellet reasoner in protege is used to test this ontology. It takes inferences from SWRL rules and outputs the results based on the patient's information, as illustrated in Figure 11.

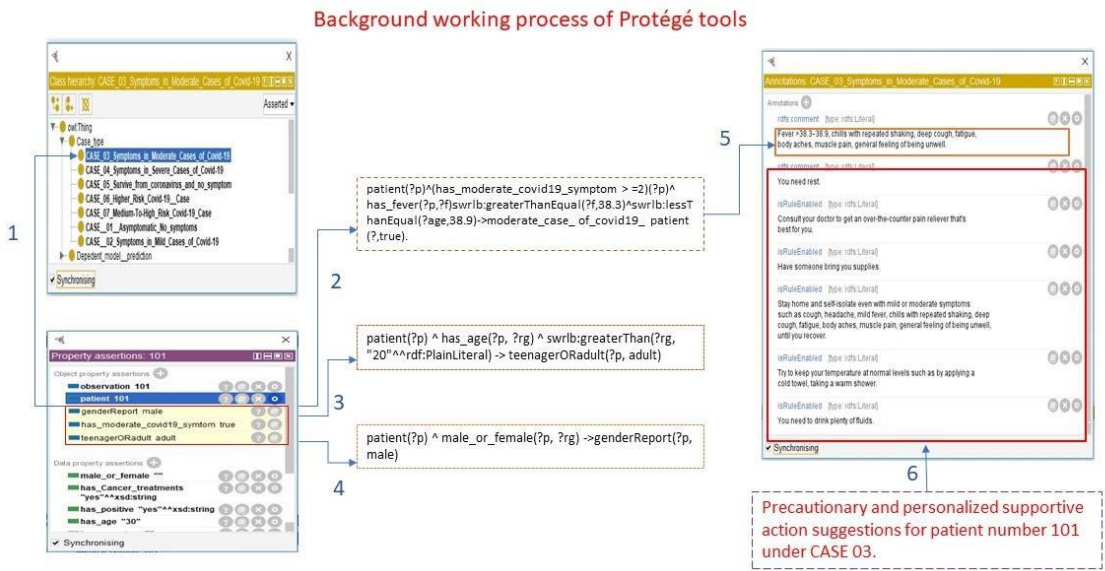

Figure 11: Working of Protege tool

Figure 11 depicts the protege tool's interface face structure, which demonstrates this reasoning.

1. Inference output gives the moderate COVID-19 case, which comes under the CASE 03 class.
2. Moderate inference rules retrieve the suggestions in accordance with the parameters, which is revealed in step 5 of the process.
3. SWRL regulations determine whether a patient is an adult or a minor.
4. The SWRL rules are required to verify the patient's gender.
5. Take a look at the patient's parameter information, which is depicted in figure 11 on the right side with a yellow outline.
6. Figure 12 illustrates which parameters should be checked in moderate cases, such as fever, cough, and body aches.

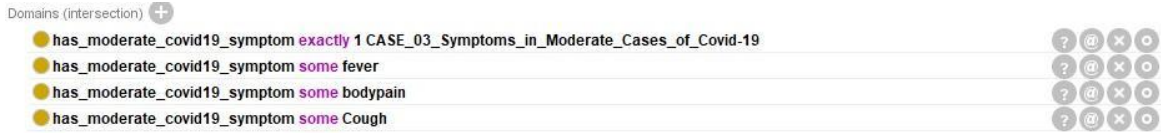

Figure 12. Moderate Case 03 Depends on Cough, fever and body pain

7. Finally, it informs you of all precautionary measures (https://www.who.int/emergencies/diseases/novel-coronavirus-2019/advice-for-public) as well as what to do in the current circumstances, based on recommendations for action. We can utilize SPARQL Query, which is covered in a subsequent part, to learn more about a patient.

Figure 13, tells about execution flow of patient 101 after running the reasoner using pellets and takes 196 ms time for the execution. Four explanations for each output: "genderReport" is "true", "has_moderate_COVID_19_symtoms" is "true" and "teenagerORadult" is "adult". Each explanation is linked with precautionary suggestions for COVID-19. This explanation is shown by the interface face framework of the protege tool which is shown in Figure 11.

Figure. 13. Patient 101 execution flow console.

## 3.3.4 Fusion of FBprophet and ARIMA models in designing COVID-19 Ontology for predicting COVID-19 Cases

We examined the ARIMA and FBprophet models in the preceding section. Now we will explain how we used time series data to create inference rules that made our ontology more perfect. We get the result with a time series which helps in taking precautionary measures on time. It also helps to find out the average death rate for every place using SWRL rules based on time, since weekly deaths, biweekly deaths, new deaths, and total deaths are all accessible in a triplestore format [36]. Figure 14 in yellow shows the inference results after running the pellet ARIMA and FB prophet models, as well as a description of the patient's location, date, and new cases.

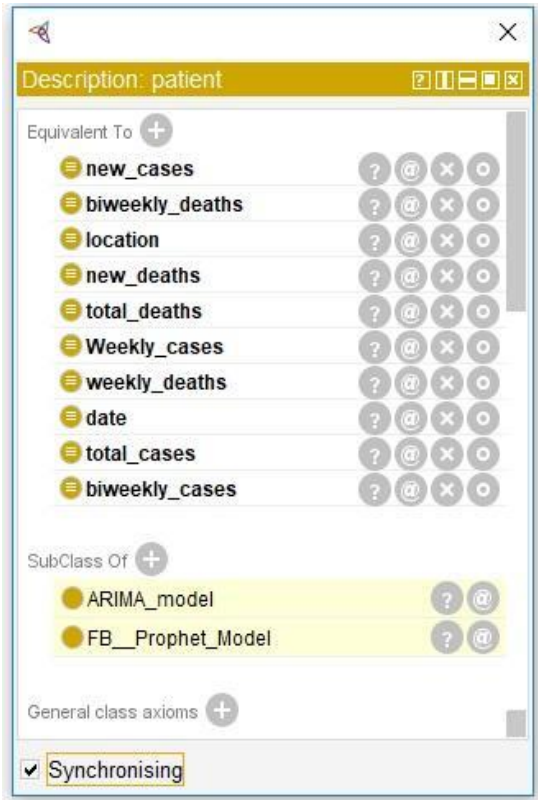

Figure. 14. ARIMA and FBprophet model applied on desperation of Patient details

Prediction model 1, Prediction model 2, ARIMA, and FP_ prophet model are all correlated in COVID-19 ontology, making it more suited and excellent. It also employs 43 SWRL rules for mathematical reasoning and ontology restrictions. Most sorts of questions are addressed by our ontology, such as (1) death-related inquiries handled by the ARIMA and FB prophet models, such as average death per week, month, and year, as well as new cases. (2) Prediction Mode 1 and 2 cover specific patient inquiries such as the likelihood of having COVID-19 or not. It also notifies about the patient's chances of survival based on their history of sickness (like diabetes, having any lung diseases etc). It reassures all the preventive steps that should be taken based on the scenario. Figure 15 depicts how the four models are connected to one another. All models that work on patient details include details about time series, gender, age, and so on. It also explains case types, which were already covered in the preceding section.

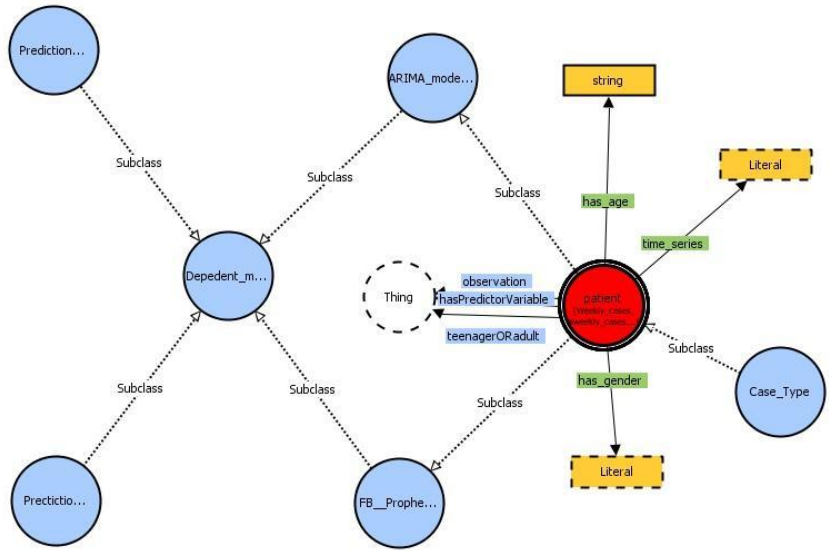

Figure 15. ProtégéVOWL visualization of ontologies [38]

### 3.3.5 Querying on Covid-19 Ontology with SPARQL Query

The SPARQL query engine allows us to query on ontologies [39]. SPARQL may be used to express searches across a range of data sources, whether the data is stored natively as RDF [36] or accessible as RDF via middleware. SPARQL may query both required and optional graph patterns, as well as their conjunctions and disjunctions. SPARQL also supports extensive value checking and query restrictions that are based on the underlying RDF graph. In addition to querying, SPARQL may delete, insert, and change data. Figure 16 shows the Protege tools output for the patient with an omicron using a SPARQL query.

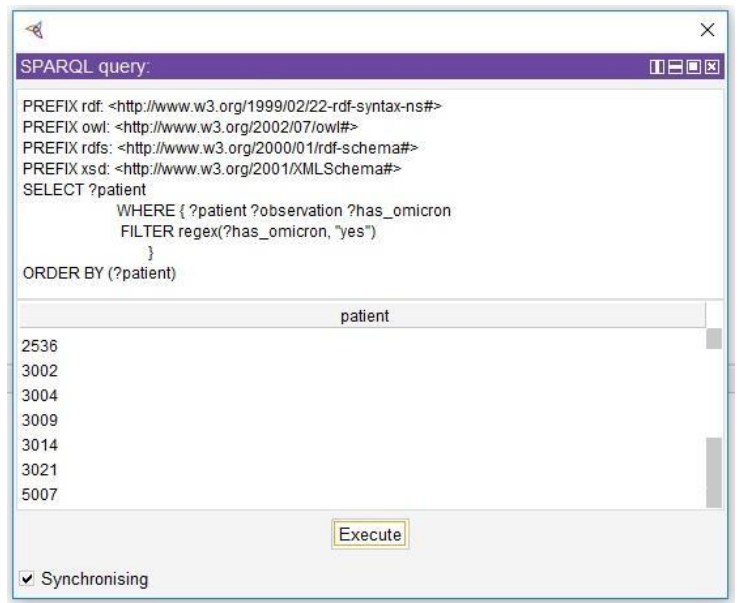

Figure 16. SPARQL results have which patients have omicrons

Also in Table 4, two other query which have finding age who greater than 30, has COVID positive, patient who have diagnosis. Another query finding each day's total active case is given in Table 5. Same way query On COVID-19 Ontology about any scenario of patient related to COVID which helps in Decision making in pandemic situations.

Table 4: SPARQL query for patient diagnosis detected COVID positive who has age greater than 30

```
PREFIX rdf: <http://www.w3.org/1999/02/22-rdf-syntax-ns#>
PREFIX owl: <http://www.w3.org/2002/07/owl#>
PREFIX rdfs: <http://www.w3.org/2000/01/rdf-schema#>
PREFIX xsd: <http://www.w3.org/2001/XMLSchema#>
SELECT DISTINCT ?patient ?gender ?age ?Diagnosis

    WHERE { ?patient rdf:type ?has_positive.
        ?patient owl:has_age ?age.
        ?patient owl:has_gender ?gender.
        ?patients rdf: is_diagnose ?Diagnosis.
         FILTER(?age>30)}
         ORDER BY (?Patient)
```

Table 5: SPARQL query for each day's total active case

```
PREFIXrdf:<http://www.w3.org/1999/02/22-rdf-syntax-ns#>
PREFIX owl: <http://www.w3.org/2002/07/owl#>
PREFIX rdfs: <http://www.w3.org/2000/01/rdf-schema#>
PREFIX xsd: <http://www.w3.org/2001/XMLSchema#>
SELECT ?datePosted (sum(?totalActiveCase) as ?nationalTotal)

FROM <http://www.w3.org/1999/02/22-rdf-syntax-ns#>
        {
     ?prov rdf :Covid_Case_In_India_Karnatka ;
        rdf:totalActiveCase ?totalActiveCase ;
         xsd:datePosted ?datePosted.
        }
GROUP BY ?datePosted
ORDER BY desc(?datePosted)
```

## 4. Experimental Results and Discussion

In this section, the proposed model uses statistical prediction models, ARIMA and FBProphet, to forecast the upcoming cases. ARIMA can be used for prediction if the data is stationary. Therefore, various techniques have been applied to check the stationarity of data and transform the data if it is not stationary. In this experiment, data stationarity is checked using an augmented dicky-fuller and for deciding the optimal values of 'p' and 'q', PACF and ACF plots are used in the case of SARIMA model performance. The FBProphet model can be directly applied to the original dataset. The accuracy results are computed using both the prediction models for active COVID cases in India. The evaluation metrics and order used for the SARIMA model are mentioned in table 6.

Table 6: Evaluation Metrics for SARIMA Model

| | |
|---|---|
| RMSE | 1126.1109 |
| MSE | 1268125.8077 |
| MAE | 862.85 |

```
new_cases    44896.4
dtype: float64
The RMSE value is:
1126.1109216076215
The MSE value is:
1268125.807763967
The MAE value is:
862.852441312852
```

The evaluation metrics computed using FBProphet model are given in table 7.

Table 7: Evaluation Metrics for FB Prophet Model

| | horizon | mse | rmse | mae | mape | mdape | coverage |
|---|---|---|---|---|---|---|---|
| 0 | 1 days 12:00:00 | 6.127777e+07 | 7828.011870 | 5502.866359 | 0.112011 | 0.092123 | 0.325000 |
| 1 | 2 days 00:00:00 | 6.206008e+07 | 7877.822144 | 5227.063354 | 0.120038 | 0.089970 | 0.341071 |
| 2 | 2 days 12:00:00 | 7.669193e+07 | 8757.392627 | 6223.874417 | 0.132739 | 0.118373 | 0.300000 |
| 3 | 3 days 00:00:00 | 7.167713e+07 | 8466.234547 | 5826.624721 | 0.135411 | 0.129228 | 0.267857 |
| 4 | 3 days 12:00:00 | 7.676805e+07 | 8761.737572 | 6141.401291 | 0.127829 | 0.101415 | 0.300000 |

The experimental results show the identified ARIMA order fits better and improves the performance due to the lower MAE, RMSE values.

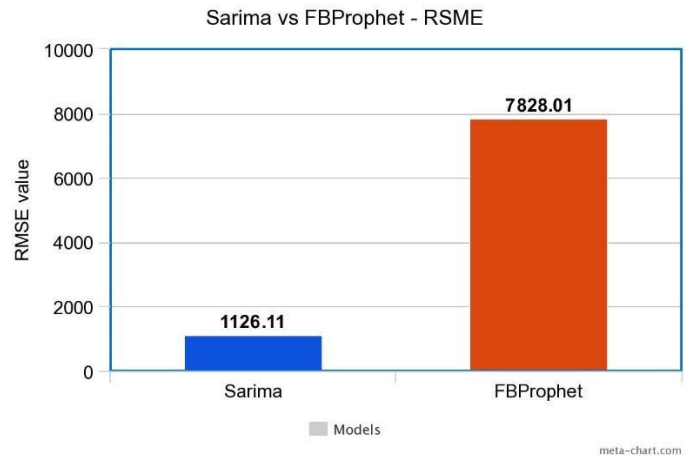

Figure 17. Comparative analysis of RMSE values

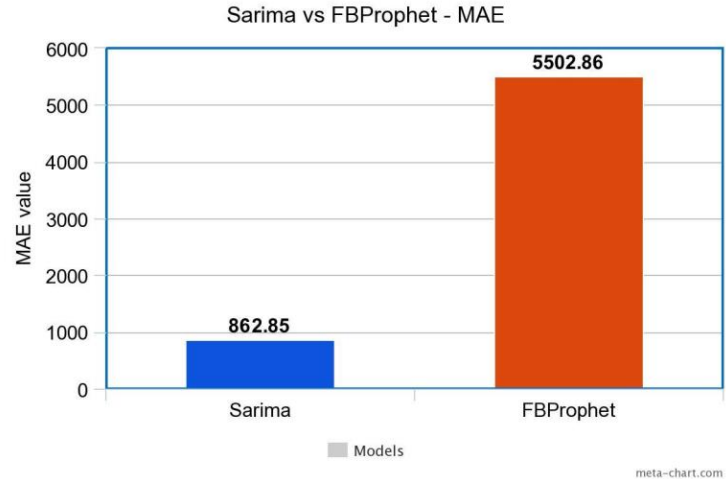

Figure 18: Comparative analysis of MSE values

Figure 17 and figure 18 show that the prediction errors using SARIMA are very small as compared to the FBProphet model, which has a high error. It is observed that the SARIMA model performs better for the correct prediction of infected, recovered, and death cases to provide proper movement of the services during the COVID pandemic situation. The estimated range of COVID-19 figures reported by the SARIMA model and FBProphet models for the upcoming days during the first wave of the pandemic can be visualized through figure 19 and 20. By observing the graph mentioned in figure 19 and 20, it can determine that hereafter the risk of contamination in India is likely to remain constant for a while and eventually decrease. The speculated overall recovery rate in India holds an optimistic trend for the upcoming days as the recovery rate is considerably higher than the number of new infections.

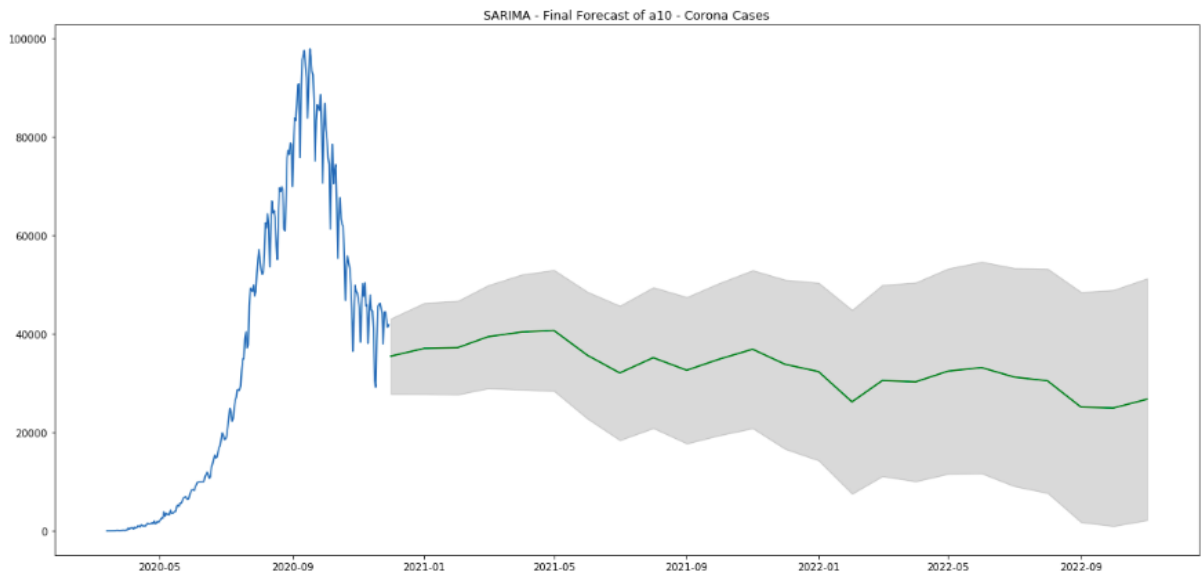

Figure 19: Forecast Results of ARIMA Model

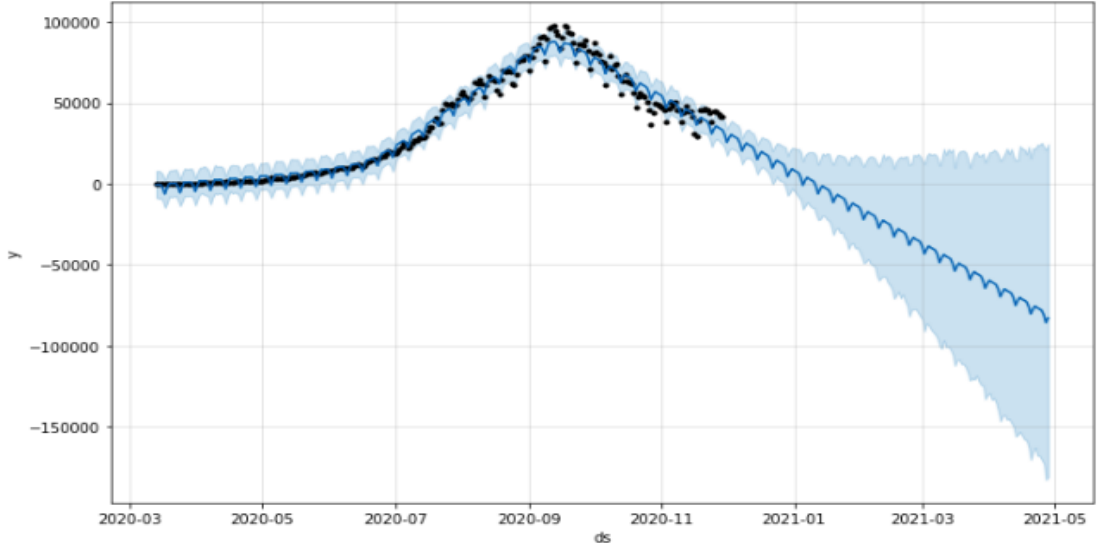

Figure 20: Forecast Results of FB Prophet Model

Table 8 shows successful test cases of COVID patients in different complications like having COVID, no COVID etc. and computes the accuracy of every complication category and figure 21 analyses the same by doing comparative analysis of five defined classes as per complexity levels.

Table 8: Test cases performed through COVID-19 Ontology

| S. No. | Complications | No. of test cases | No. of Successful test cases | Accuracy (%) |
|---|---|---|---|---|
| 1 | Having COVID | 89 | 74 | 83 |
| 2 | No COVID | 94 | 87 | 92 |
| 3 | High Risk COVID case | 74 | 63 | 85 |
| 4 | Medium to High Risk case | 102 | 83 | 81 |
| 5 | Control Needed case | 98 | 90 | 91 |
|  |  | 457 | 397 |  |

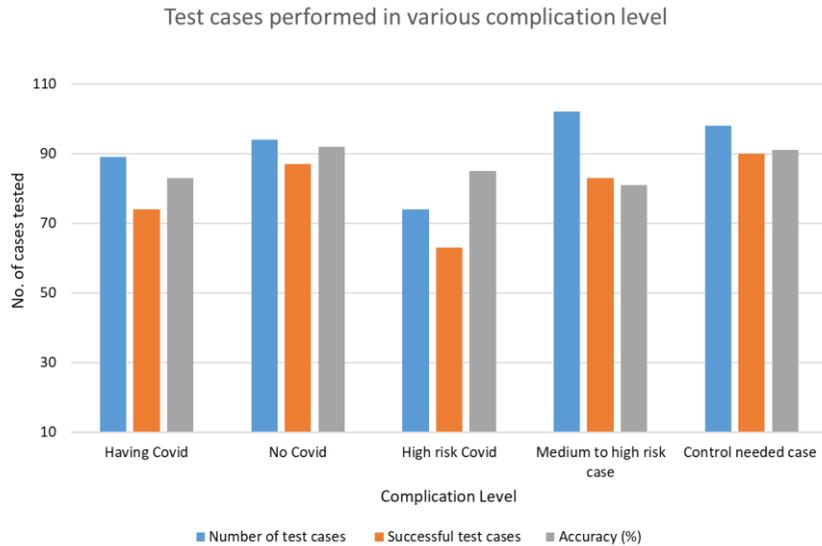

Also check the ontology quality score through the metric of the ontology [40] based on the knowledge being represented in figure 22, which reflects the relationship and qualities of the domain knowledge according to equation 3.

$$\text{Score\_rk} = \frac{(|rel|*|class|*100) + (|sub-class|+|rel|)*|prop|}{(|sub-class| + |rel|)*|class|} \ldots\ldots\ldots\ldots (3)$$

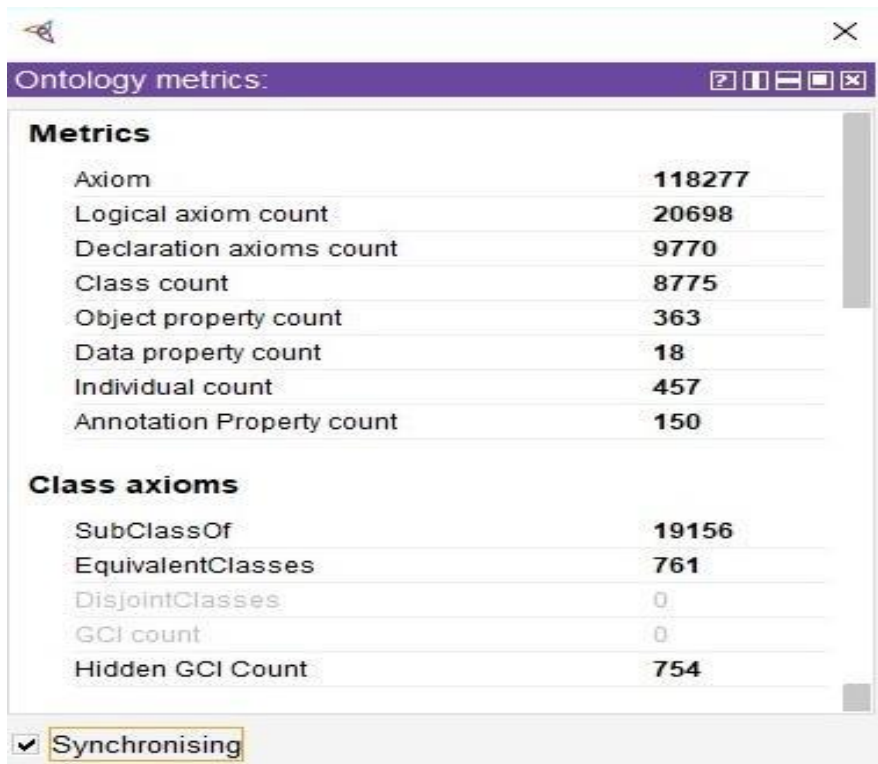

Figure 22. Metrics of COVID-19 ontology

On the basis of the efficiency with which base knowledge can be extracted [40], according to equation 4.

$$\text{Score\_bk} = \frac{(Class\ with\_instance * 100) + (|Individual|)}{|class|} \ldots\ldots\ldots\ldots\ldots (4)$$

COVID 19 Ontology score is based on the formula in table 9 for score_rk and score_bk.

Table 9: COVID 19 Ontology Score

| Evaluation parameter | Score |
|---|---|
| Score_rk | 92.07 |
| Score_bk | 96.02 |

## 5. Conclusion and Future Scope

This paper explores the relevance and usefulness of various time series prediction models for analysing COVID-19 cases in India. A SARIMA-based prediction model, and FBProphet model are used for the prediction of the total number of daily new cases. The proposed model is fruitful in the current scenario for predicting future infected cases if the virus spread pattern does not change adversely. When the SARIMA model is used in the COVID-19 scenario, the predictions are generated on the basis of the series' prior values and error lags, which allows the model to alter its forecast values in the event of a sudden shift in trends. Since our interest is in creating short-term predictions using time series data and generating the outcome in the coming months, the SARIMA model appeared to be the best fit. A COVID - 19 ontology is also designed to perform individual patient prediction into five designated classes such as Having COVID, No COVID, High Risk COVID case, Medium to High Risk case, and Control needed case. Apart from this, ontology based semantic modelling helps in predicting the chances of suffering from severe coronavirus infection depending on various individual factors like age, weight, gender, immune system, and having previously suffered from any other chronic disease. It helps in a better way to calculate the percentage score of getting affected by the virus and suggests what precautions and treatment should be taken while suffering from this infectious disease. Although this research work was performed on the SARS-Cov-2 dataset, which is available on Kaggle, in the future, the proposed model can be developed for the identification of Deltacron variant cases and more SWRL rules can be constructed across future situations based on Deltacron. So semantic models work more accurately according to the future development.

### Acknowledgements

This research is supported by "Extra Mural Research (EMR) Government of India Fund by Council of Scientific & Industrial Research (CSIR)", Sanction letter no. – 60(0120)/19/EMR-II. Authors are thankful to CSIR for providing the necessary equipment for the research. Authors are also thankful to the authorities of "Indian Institute of Information Technology, Allahabad at Prayagraj ", for providing us infrastructure and necessary support.

## Author's Bibliography

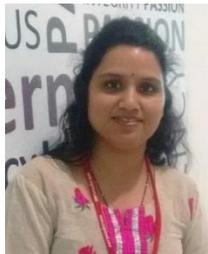 Sadhana Tiwari received the B. Tech. Degree in Computer with a focus in CSE and completed the M. Tech. Degree in Software Engineering from IIIT Allahabad, India. Currently I am working as a research scholar in IIIT Allahabad, India in the area of big data analysis and stream data mining in biomedical domain.

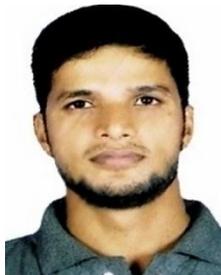 Ritesh Chandra received the B. Tech. Degree in Computer with a focus in CSE and completed the M. Tech. Degree in Information Technology from NIT Patna, India. Currently I am working as a research scholar in IIIT Allahabad, India in the area of Ontology Engineering and big data analysis in healthcare domain.

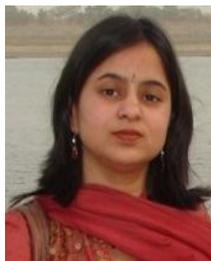Sonali Agarwal, working as an Associate Professor in the Information Technology Department of the Indian Institute of Information Technology (IIIT), Allahabad, India. She received her Ph. D. Degree at IIIT Allahabad and joined as faculty at IIIT Allahabad since October 2009. The main research interests are in the areas of Stream Analytics, Big Data, Stream Data Mining, Complex Event Processing System, Support Vector Machines and Software Engineering. Dr. Sonali Conducted many workshops and seminars related to the area of big data analytics and real time stream analytics. She received many funded projects form DST, SERB and CSIR etc.